\documentclass[journal]{IEEEtran}

\newcommand{\papertitle}{Low-Frequency Magnetic Noise in Statically-Driven Solenoid for Biasing Magnetic Field Sensors}

\usepackage{cite}

\ifCLASSINFOpdf
  \usepackage[pdftex]{graphicx}
\else
\fi
\usepackage{amsmath}
\usepackage{sistyle}

\usepackage{hyperref}
\hypersetup{
    unicode=false,
    pdftoolbar=true,
    pdfmenubar=true,
    pdffitwindow=false,
    pdfstartview={FitH},
    pdftitle={\papertitle},
    pdfauthor={Durdaut et al.},
    pdfsubject={\papertitle},
    pdfcreator={Durdaut et al.},
    pdfproducer={Kiel University},
    pdfnewwindow=true,
    colorlinks=true,
    linkcolor=black,
    citecolor=black,
    filecolor=black,
    urlcolor=black
}

\usepackage{graphicx}
\usepackage{caption}
\usepackage{subcaption}

\usepackage{epstopdf}

\usepackage{footnote}
\usepackage[flushleft]{threeparttable}

\usepackage{upgreek}

\usepackage{url}

\usepackage{cancel}

\usepackage{booktabs}

\usepackage{multirow}

\usepackage{mathtools}

\usepackage{cuted}
\usepackage{mathrsfs}

\usepackage{listings}
\usepackage{color}

\definecolor{dkgreen}{rgb}{0,0.6,0}
\definecolor{gray}{rgb}{0.5,0.5,0.5}
\definecolor{mauve}{rgb}{0.58,0,0.82}

\usepackage{array}
\begin{document}
\bstctlcite{IEEEexample:BSTcontrol}

\title{\papertitle}

\author{Phillip~Durdaut,
				Henrik~Wolframm, 
        and Michael~H\"oft
\thanks{P.~Durdaut, H.~Wolframm, and M.~H\"oft are with the Chair of Microwave Engineering, Institute of Electrical Engineering and Information Technology, Faculty of Engineering, Kiel University, Kaiserstr. 2, 24143 Kiel, Germany.}}

%
%

\markboth{}{}
%



\maketitle


%
\IEEEpeerreviewmaketitle

\begin{abstract}
For the generation of static magnetic fields solenoids are frequently used for the purpose of research and development of magnetic field sensors. When such a sensor is to be analyzed with regard to its inherent noise the influence of other noise sources of the measurement system needs to be minimized. This article reports on the low-frequency magnetic noise within such a coil system. On the one hand, the impact of the intrinsic noise of the coil itself and on the other hand the impact of additional current noise of various commercially available current sources, which accordingly also leads to magnetic noise within the coil, are investigated. With low-frequency values in the range of a few tens of $\mathrm{fT}/\sqrt{\mathrm{Hz}}$, the coil's inherent noise is mostly neglectable. However, frequently utilized current sources for the generation of a static magnetic bias field lead to significant low-frequency magnetic flux noise typically in the $\mathrm{nT}/\sqrt{\mathrm{Hz}}$ regime. It is found that this noise cannot be decreased by increasing the coil's magnetic sensitivity, i.e. the magnetic flux density as a function of the static current. Instead, current sources with very high current-to-current-noise ratios are required.
\end{abstract}

\section{Introduction}

Similar to biasing bipolar \cite[pp. 83-88]{Hor15} or field-effect transistors \cite[p. 149]{Hor15} by static currents and voltages, respectively, various types of magnetic field sensors also require a magnetic biasing for maximizing their sensitivity, e.g. sensors based on the magnetostrictive effect. For practical applications, a suitable operating point can be realized via magnetostrictive multilayers, i.e. an exchange bias system \cite{Lag12}. For the purpose of research and development electromagnets are generally used to determine field-dependent properties of a sensor. With regard to the characterization of signal properties or sensitivities coils are generally well suited. However, if the intrinsic noise of a magnetic field sensor is to be determined as a function of an external and static magnetic field, see e.g. \cite{Dur20}, the influence of other noise sources of the measurement system needs to be minimized or reduced below the inherent noise of the sensor, respectively. State-of-the-art magnetic field sensors, e.g. sensors based on the $\Delta$E effect, currently reach inherent noise levels as low as e.g. \SI{70}{pT/\sqrt{Hz}} at a frequency of \SI{10}{Hz} \cite{Sch20}. Thus, the additional magnetic noise of the biasing coil system needs to be distinctly smaller than this value, e.g. \SI{1}{pT/\sqrt{Hz}} at a frequency of \SI{10}{Hz}. Apart from certainly existing exceptions, the frequency range of interest is approximately from \SI{1}{Hz} (often also below) to \SI{1}{kHz}.

This article reports on the low-frequency magnetic noise inside such a coil system. All investigations were carried out on a specific coil, the structure of which is explained below. Two different noise sources are considered. On the one hand, the impact of the intrinsic noise of the coil itself and on the other hand the impact of additional current noise of various commercially available current sources, which accordingly also leads to magnetic noise within the coil, are investigated.

\section{Coil System}

\begin{figure}[t]
	\centering
	\includegraphics[width=0.35\textwidth]{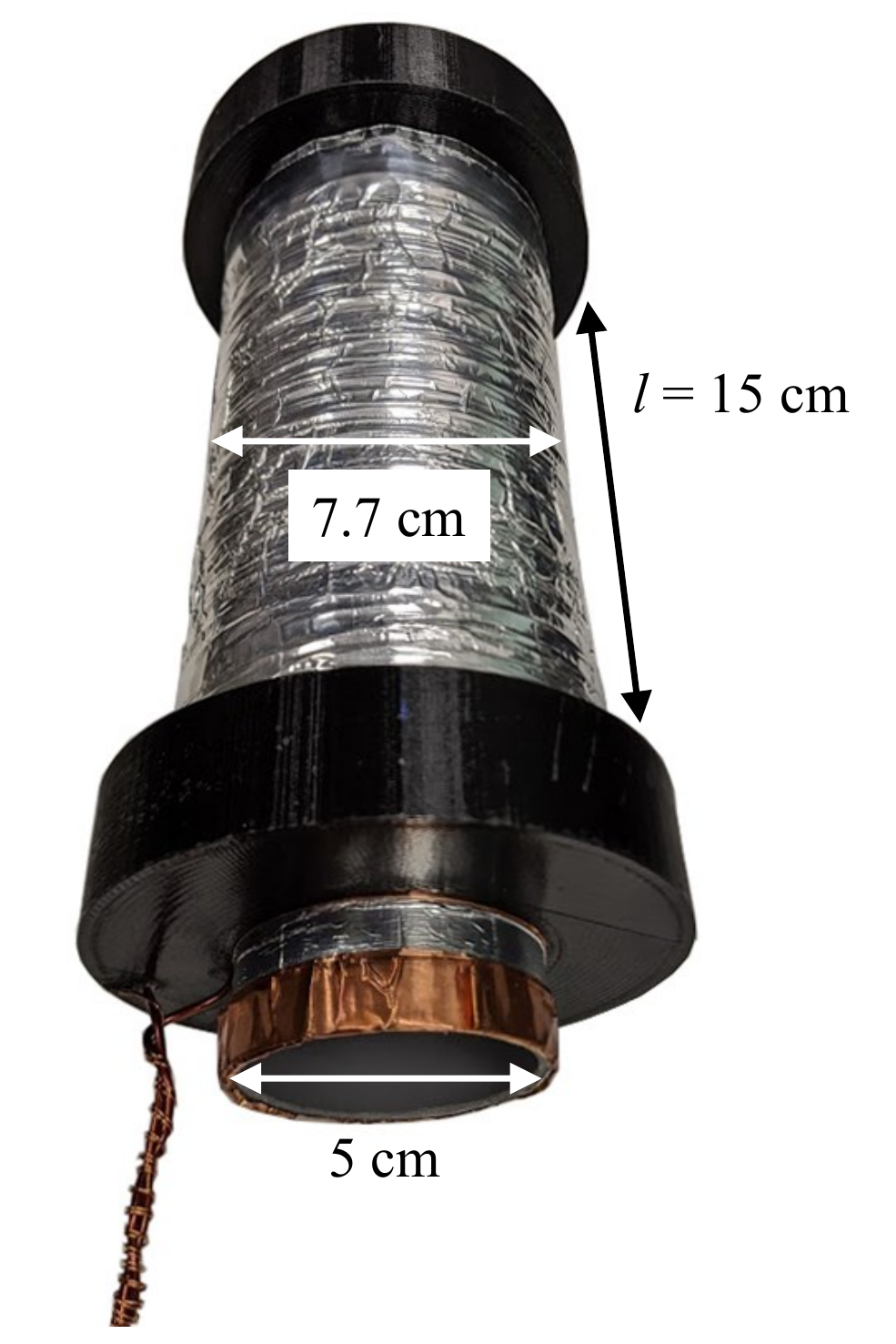}
	\caption{Photography of the coil system under investigation consisting of two coils wound on top of each other. The inner coil (\textit{DC coil}) is used for generating static bias magnetic fields and the additional outer coil (\textit{AC coil}) enables the generation of dynamic magnetic test signals.}
	\label{fig:coil1}
\end{figure}

\begin{table*}[t]
	\renewcommand{\arraystretch}{1.3}
	\caption{Overview of all relevant parameters of the investigated coil system.}
	\label{tab:coil_parameters}
	\centering
	\begin{tabular}{c | c | c | c}
		\toprule
		Parameter		 																	& Symbol																																	& Inner coil (DC coil)									& Outer coil (AC coil)\\
		\hline
		Coil length																		& $l$																																			& \SI{15}{cm}														& \SI{15}{cm}\\
		Number of turns																& $N$																																			& $8 \text{ x } 90 = 720$								& $2 \text{ x } 300 = 600$\\		
		Wire diameter																	& $d_{\mathrm{w}}$																												& \SI{1.5}{mm}													& \SI{0.4}{mm}\\
		Inner diameter																& $d_{\mathrm{i}}$																												& \SI{5}{cm}														& \SI{4.5}{cm}\\
		Outer diameter																& $d_{\mathrm{o}}$																												& \SI{7.4}{cm}													& \SI{7.7}{cm}\\
		Effective diameter														& $d_{\mathrm{eff}} = d_{\mathrm{i}} + (d_{\mathrm{o}}-d_{\mathrm{i}})/2$	& \SI{6.2}{cm}													& \SI{7.6}{cm}\\
		Cross-section area														& $A = \pi \cdot (d_{\mathrm{eff}}/2)^2$																	& \SI{30.2}{cm^2}												& \SI{45.4}{cm^2}\\
		Coil sensitivity 															& $S = \mu_0 N / l$																												& \SI{6.03}{mT/A}												& \SI{5.03}{mT/A}\\	
																									& 																																				& \SI{5.95}{mT/A} (measured)						& \SI{5.06}{mT/A} (measured)\\		
		Coil inductance 															& $L = \mu_0 N^2 A / l$																										& \SI{13.1}{mH}													& \SI{13.7}{mH}\\		
																									& 																																				& \SI{10.4}{mH}	(measured)							& \SI{10.2}{mH}	(measured)\\		
		Coil capacitance 															& $C$																																			& \SI{770}{pF} (measured)								& \SI{980}{pF} (measured)\\
		Parallel resonance frequency									& $f_{\mathrm{res}} = 1/(2 \pi \sqrt{L C})$																& \SI{56.2}{kHz} (measured)							& \SI{50.3}{kHz} (measured)\\
		Wire length																		& $l_{\mathrm{w}} = N \pi d_{\mathrm{eff}}$																& \SI{140}{m}														& \SI{143}{m}\\		
		Wire resistance per meter											& $r_{\mathrm{w}}$																												& \SI{21.8}{m\Omega/m}									& \SI{136}{m\Omega/m}\\
		Wire resistance 															& $R_{\mathrm{w}} = r_{\mathrm{w}} l_{\mathrm{w}}$												& \SI{3.1}{\Omega}											& \SI{19.5}{\Omega}\\
																									& 																																				& \SI{1.43}{\Omega}	(measured)					& \SI{19.28}{\Omega} (measured)\\
		Proximity effect loss resistance 							& $R_{\mathrm{p}}$																												& \SI{550}{\Omega} (measured)						& \SI{1.8}{k\Omega} (measured)\\
		Intrinsic magnetic noise ($f < \SI{1}{kHz}$) 	& $B_{\mathrm{n}}$																												& $32~\mathrm{fT}/\sqrt{\mathrm{Hz}}$		& $15~\mathrm{fT}/\sqrt{\mathrm{Hz}}$\\
		
		\bottomrule
	\end{tabular}
\end{table*}

The coil system under investigation of which a photography is shown in Fig.~\ref{fig:coil1} consists of two cylindrical coils, i.e. two solenoids, which are wound on top of each other. The reason for the requirement of two coils is that in measurements for the characterization of magnetic field sensors typically not only a coil for the static bias field is needed, in the following referred to as \textit{DC coil}, but also an additional coil for the generation of dynamic test fields is required, in the following referred to as \textit{AC coil}.

Both coils are wound on a DN 40 plastic pipe with an outer diameter of \SI{5}{cm}. By means of two 3D printed stoppers (black) with a diameter of \SI{10}{cm} the coil system directly fits into an ultra high magnetic field shielding mu-metal cylinder \textit{ZG1} from \textit{Aaronia AG} in which all magnetic measurements are typically performed. The inner distance between the two stoppers yields the length of both coils of ${l = \SI{15}{cm}}$. The inner coil (DC coil) consists of 8 layers of 90 windings each, resulting in a total number of turns of ${N = 720}$. The copper wire itself has a diameter of ${d_{\mathrm{w}} = \SI{1.5}{mm}}$ (manufacturer: Elosal). The outer coil (AC coil) consists of 2 layers of 300 windings each, resulting in a total number of turns of ${N = 600}$. The utilized copper wire has a diameter of ${d_{\mathrm{w}} = \SI{0.4}{mm}}$ (manufacturer: RS PRO). All relevant parameters of the two coils, some of which will be introduced and further explained below, are summarized in Tab.~\ref{tab:coil_parameters}.

\section{Magnetic Sensitivity}

\begin{figure}[t]
	\centering
	\includegraphics[width=0.5\textwidth]{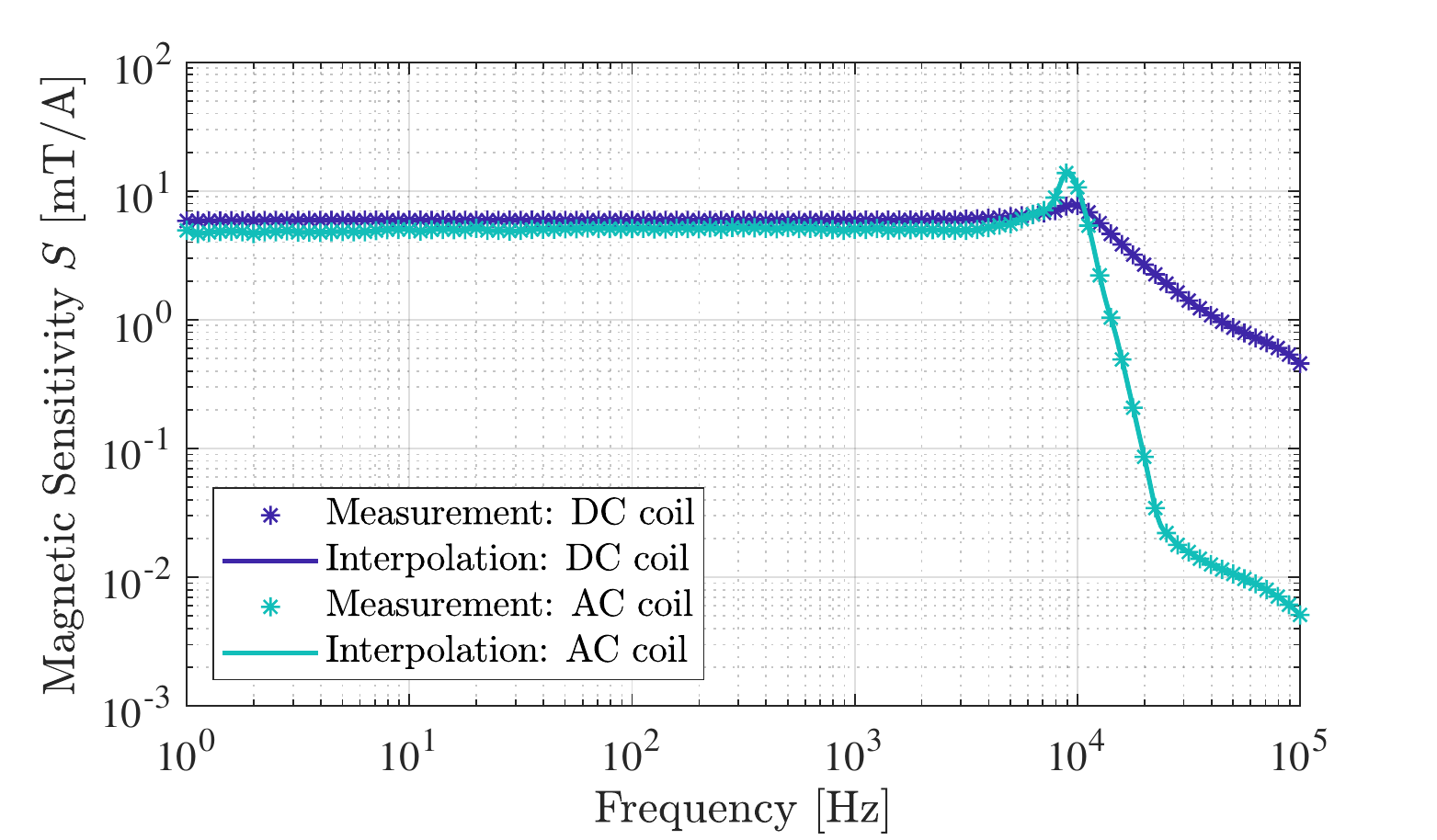}
	\caption{Measured sensitivities of the two coils as a function of the frequency revealing constant values of \SI{5.95}{mT/A} (DC coil) and \SI{5.06}{mT/A} (AC coil) for frequencies below several kilohertz.}
	\label{fig:sensitivity}
\end{figure}

One of the most important properties of the coils is their magnetic sensitivity, i.e. the generated homogenous magnetic flux density $B$ as a function of the current through the coil $I$. For a cylindrical and long air coil (relative permeability $\mu_{\mathrm{r}} = 1$) without any ferromagnetic core, the sensitivity
\begin{align}
	S = \frac{B}{I} = \frac{\mu_0 N}{l}
	\label{eq:S}
\end{align}
can be calculated based on the coil's total number of turns $N$ and its length $l$ \cite[p. 939]{Ser04}. The additional term $\mu_0$ represents the vacuum permeability (${\mu_0 = 4 \pi \cdot 10^{-7}~\mathrm{Vs}/\mathrm{Am}}$). With Eq.~\eqref{eq:S} values of \SI{6.03}{mT/A} (DC coil) and \SI{5.03}{mT/A} (AC coil) result.

Both sensitivity values were confirmed by measuring the magnetic flux density $B$ inside the coils utilizing an \textit{FM 302} teslameter (uniaxial probe \textit{AS-LAP}) from \textit{Projekt Elektronik GmbH} connected to a lock-in amplifier \textit{SR830} from \textit{Stanford Research Systems} for each coil being fed by a \textit{Keithley 6221} current source with an alternating current of an amplitude of ${\hat{I} = \sqrt{2}~I = \SI{100}{mA}}$. The results are shown in Fig.~\ref{fig:sensitivity} and reveal constant sensitivities of \SI{5.95}{mT/A} (DC coil) and \SI{5.06}{mT/A} (AC coil) for frequencies below several kilohertz. At higher frequencies above the inductor's parallel resonance frequency the coils no longer behave inductive but capacitive \cite[p. 15]{Bow97} (see the following section) leading to decreasing sensitivities.

\section{Electrical Impedance}

\begin{figure}[t]
	\centering
	\includegraphics[width=0.25\textwidth]{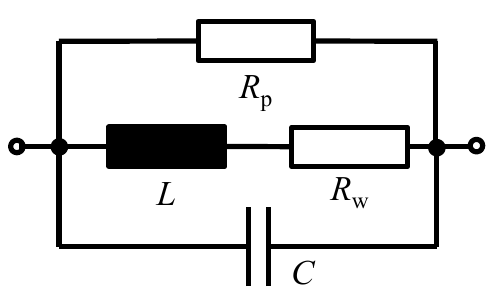}
	\caption{Equivalent circuit of an air solenoid with the inductance $L$, the wire resistance $R_{\mathrm{w}}$, and the parasitic capacitance between the windings $C$. The parallel resistance $R_{\mathrm{p}}$ covers for additional loss mechanisms.}
	\label{fig:equivalent_circuit}
\end{figure}

\begin{figure*}[t!]
\captionsetup[subfigure]{justification=centering}
	\begin{subfigure}[t]{0.5\textwidth}
		\centering
		\includegraphics[width=1\linewidth]{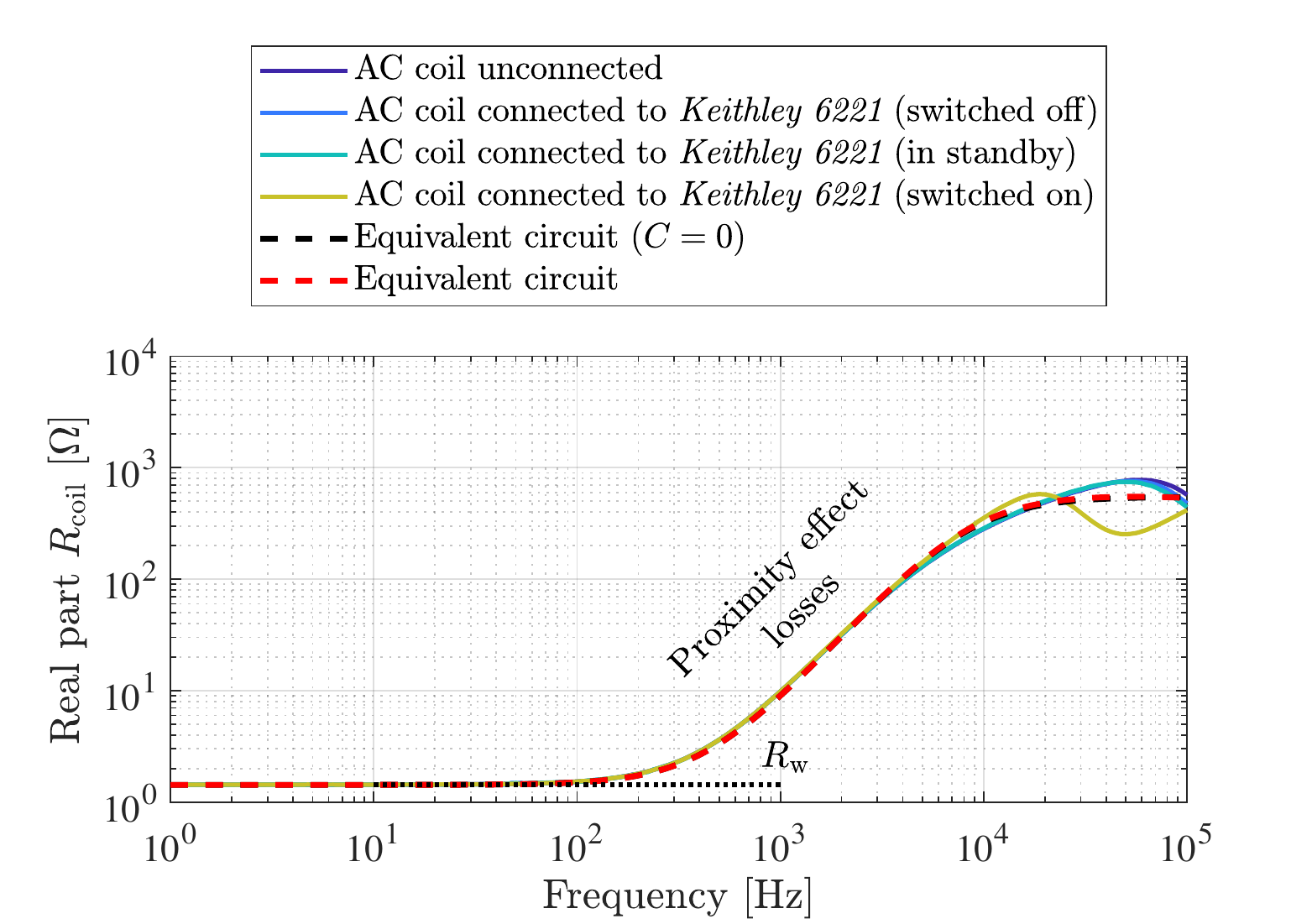}
		\caption{Real part of the DC coil's impedance}
		\label{fig:coil_impedance_r_dc}
	\end{subfigure}
	\begin{subfigure}[t]{0.5\textwidth}
		\centering
		\includegraphics[width=1\linewidth]{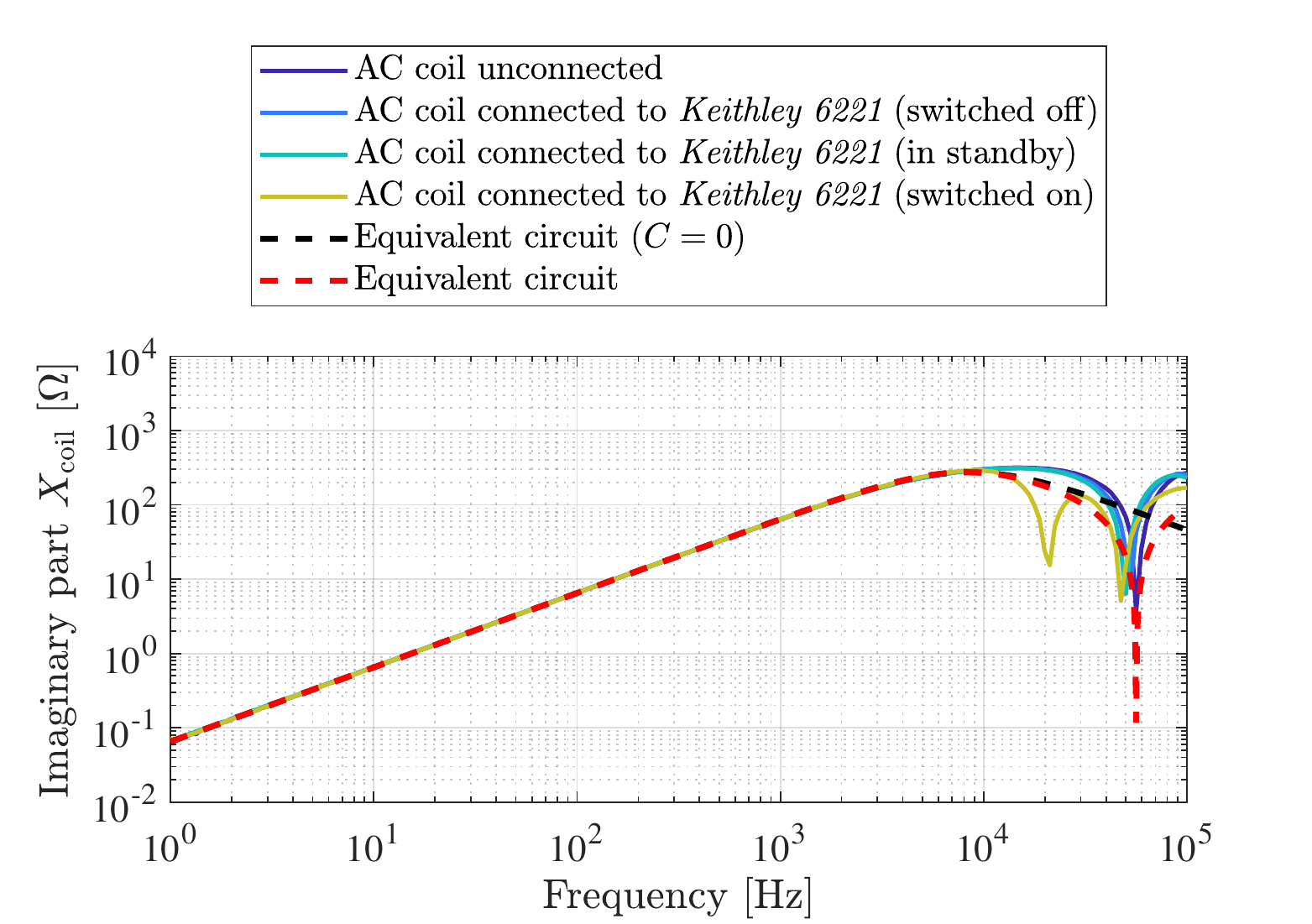}
		\caption{Imaginary part of the DC coil's impedance}
		\label{fig:coil_impedance_x_dc}
	\end{subfigure}
	\begin{subfigure}[t]{0.5\textwidth}
		\centering
		\includegraphics[width=1\linewidth]{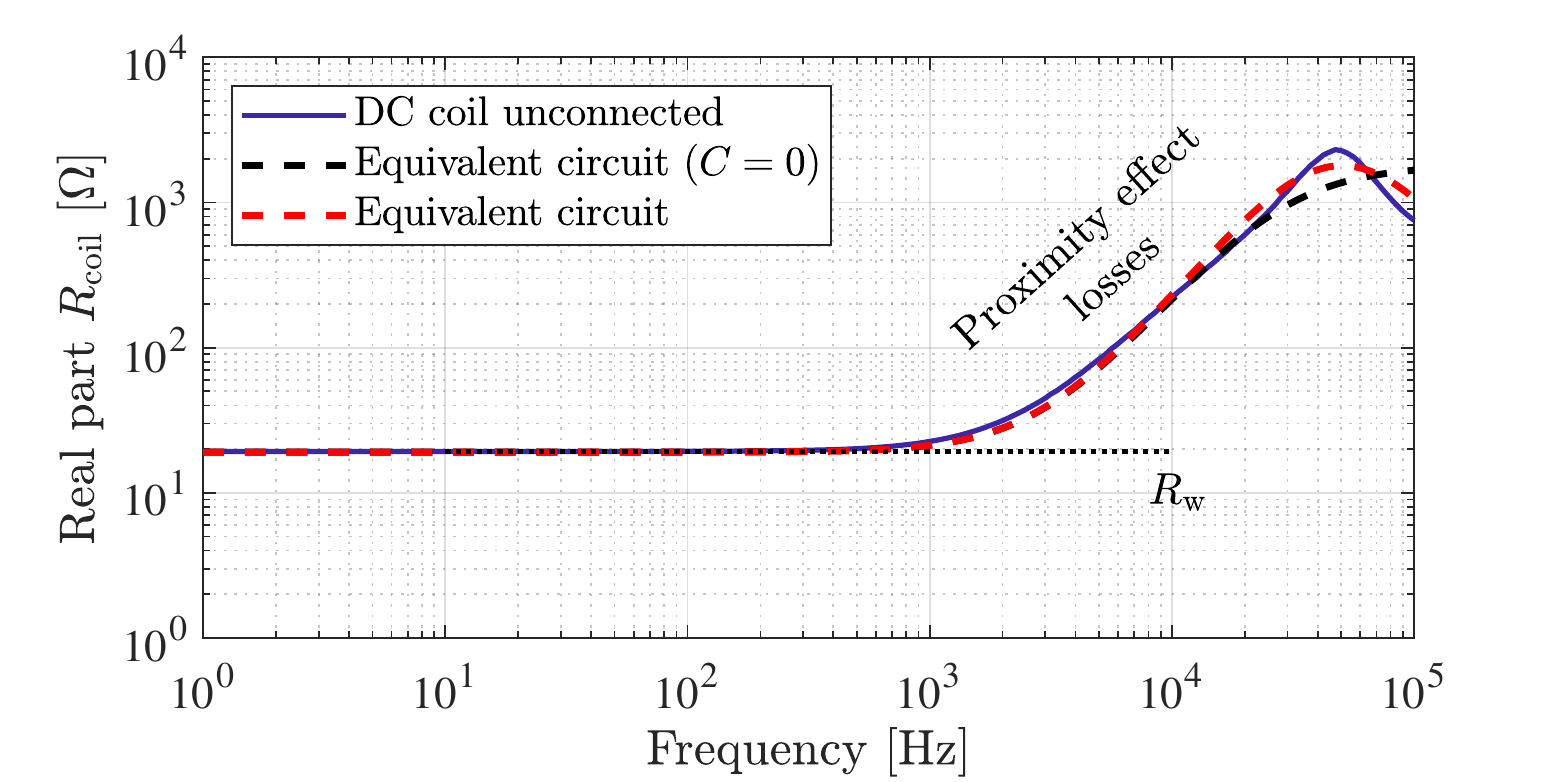}
		\caption{Real part of the AC coil's impedance}
		\label{fig:coil_impedance_r_ac}
	\end{subfigure}
	\begin{subfigure}[t]{0.5\textwidth}
		\centering
		\includegraphics[width=1\linewidth]{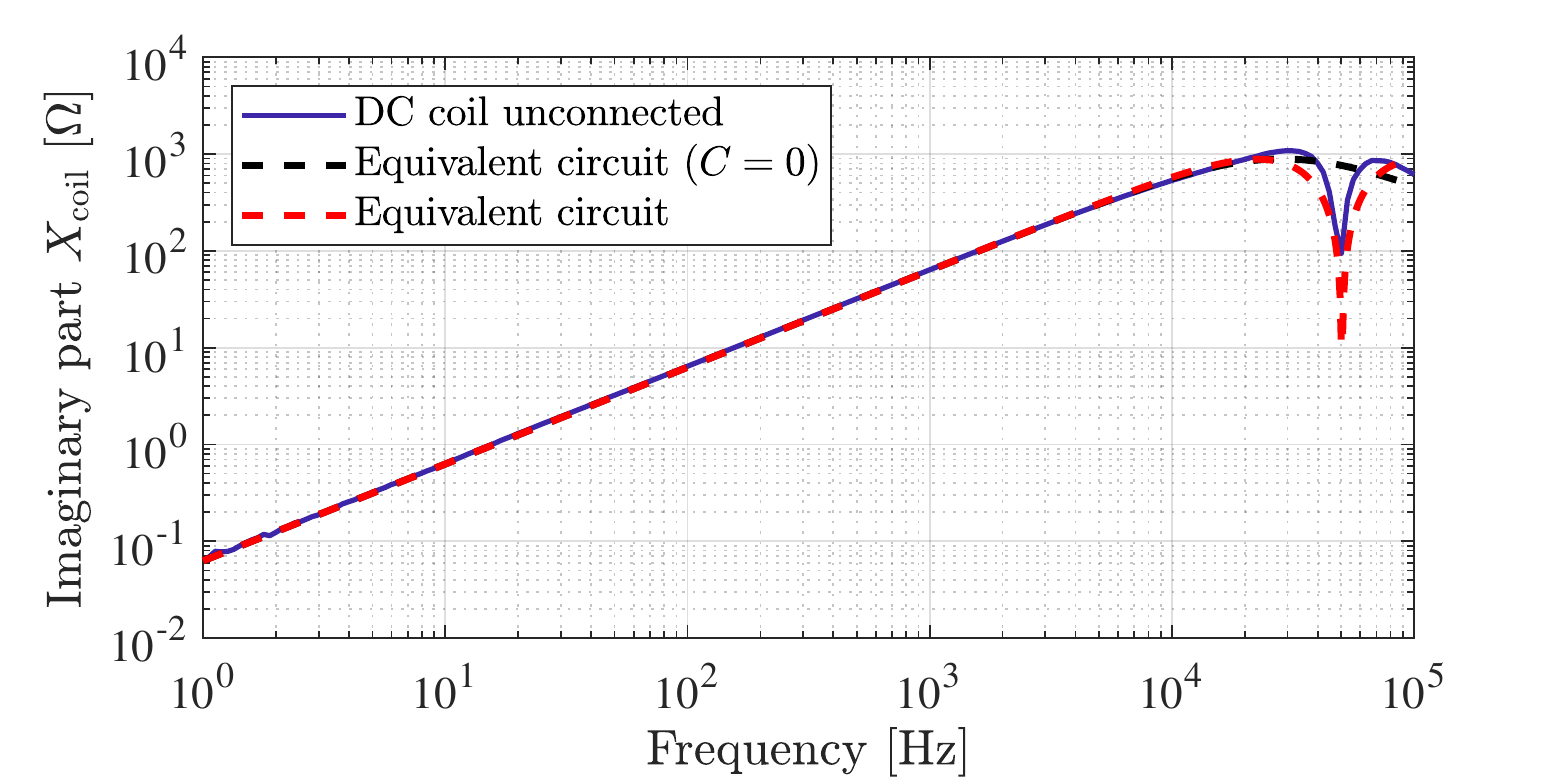}
		\caption{Imaginary part of the AC coil's impedance}
		\label{fig:coil_impedance_x_ac}
	\end{subfigure}
	\caption{Measured real (left) and imaginary (right) parts of the DC (top) and AC coil (bottom) from which the parameters of the equivalent circuit (Fig.~\ref{fig:equivalent_circuit}) are extracted as summarized in Tab.~\ref{tab:coil_parameters}. The DC coil was characterized for several connection states of the AC coil revealing that the relevant low-frequency behavior below \SI{10}{kHz} is not affected by the AC coil's connection state.}
	\label{fig:coil_impedance}
\end{figure*}

The electrical impedance ${Z_{\mathrm{coil}} = R_{\mathrm{coil}} + j X_{\mathrm{coil}}}$ of a coil is frequently described by an equivalent circuit as depicted in Fig.~\ref{fig:equivalent_circuit} \cite[pp. 15-20]{Bow97}. In this circuit, $L$ is the coil's inductance which can be calculated by
\begin{align}
	L = \frac{\mu_0 N^2 A}{l}
	\label{eq:L}
\end{align}
where $A$ is the cross-sectional area of the air solenoid \cite[p. 1006]{Ser04}. For the two coils under investigation with parameters as listed in Tab.~\ref{tab:coil_parameters}, Eq.~\eqref{eq:L} yields theoretical inductances of \SI{13.1}{mH} (DC coil) and \SI{13.7}{mH} (AC coil). Additional parameters of the equivalent circuit are the resistance of the wire $R_{\mathrm{w}}$, the parasitic capacitance between the windings $C$, and the parallel resistance $R_{\mathrm{p}}$ covering for additional losses such as losses in the coil's core material (such hysteresis losses do not occur in an air solenoid), losses due to the skin effect, losses due to eddy-currents, and losses due to the proximity effect \cite{Lot92}. 

The electrical impedance ${Z_{\mathrm{coil}} = R_{\mathrm{coil}} + j X_{\mathrm{coil}}}$ of both coils was measured utilizing a calibrated impedance analyzer \textit{Bode 100} from \textit{OMICRON Lab} with the results shown in Fig.~\ref{fig:coil_impedance}. For research and development purposes the DC coil is generally used more frequently compared to the AC coil. For this reason and because of the coupling of both coils, the impedance of the DC coil was measured for various cases. As shown in Fig.~\ref{fig:coil_impedance_r_dc} and Fig.~\ref{fig:coil_impedance_x_dc} both the real and the imaginary parts slightly depend on the connection state of the AC coil, i.e. whether it is unconnected or connected to a current source (here it is connected exemplary to a frequently utilized current source \textit{Keithley 6221}). However, these differences only occur in the frequency regime of the inductor's parallel resonance. In the practical relevant frequency range below \SI{10}{kHz} the electrical impedance of the DC coil is not significantly affected by the connection state of the AC coil. Characteristic for the measured impedance curves are the constant real part at low frequencies due to the resistance of the wire ($R_{\mathrm{w}}$) and the parallel resonance frequency \cite[p. 16]{Bow97}
\begin{align}
	f_{\mathrm{res}} = \frac{1}{2 \pi \sqrt{L C}}
	\label{eq:fres}
\end{align}
at ${f_{\mathrm{res}} = \SI{56.2}{kHz}}$ (DC coil) and ${f_{\mathrm{res}} = \SI{50.3}{kHz}}$ (AC coil), respectively. In addition, in contrast to an ideal coil, it is noticeable that the real part of the impedance increases with higher frequencies. These losses are attributed to the proximity effect and can be calculated using the \textit{Dowell} model \cite{Dow66}.

Based on the measurements of the electrical impedances all parameters of the equivalent circuit (Fig.~\ref{fig:equivalent_circuit}) can be determined, yielding values of ${L = \SI{10.4}{mH}}$, ${R_{\mathrm{w}} = \SI{1.43}{\Omega}}$, ${C = \SI{770}{pF}}$, and ${R_{\mathrm{p}} = \SI{550}{\Omega}}$ (DC coil) and ${L = \SI{10.2}{mH}}$, ${R_{\mathrm{w}} = \SI{19.28}{\Omega}}$, ${C = \SI{980}{pF}}$, and ${R_{\mathrm{p}} = \SI{1.8}{k\Omega}}$ (AC coil). As shown in Fig.~\ref{fig:coil_impedance}, both coils are sufficiently described by the equivalent circuit (red dashed lines) in the low-frequency range below \SI{10}{kHz}. In fact, for this frequency range, the capacitor can even be neglected ($C = 0$, black dashed lines).

\section{Magnetic Noise due to Coils Intrinsic Noise}

\begin{figure}[t]
	\centering
	\includegraphics[width=0.5\textwidth]{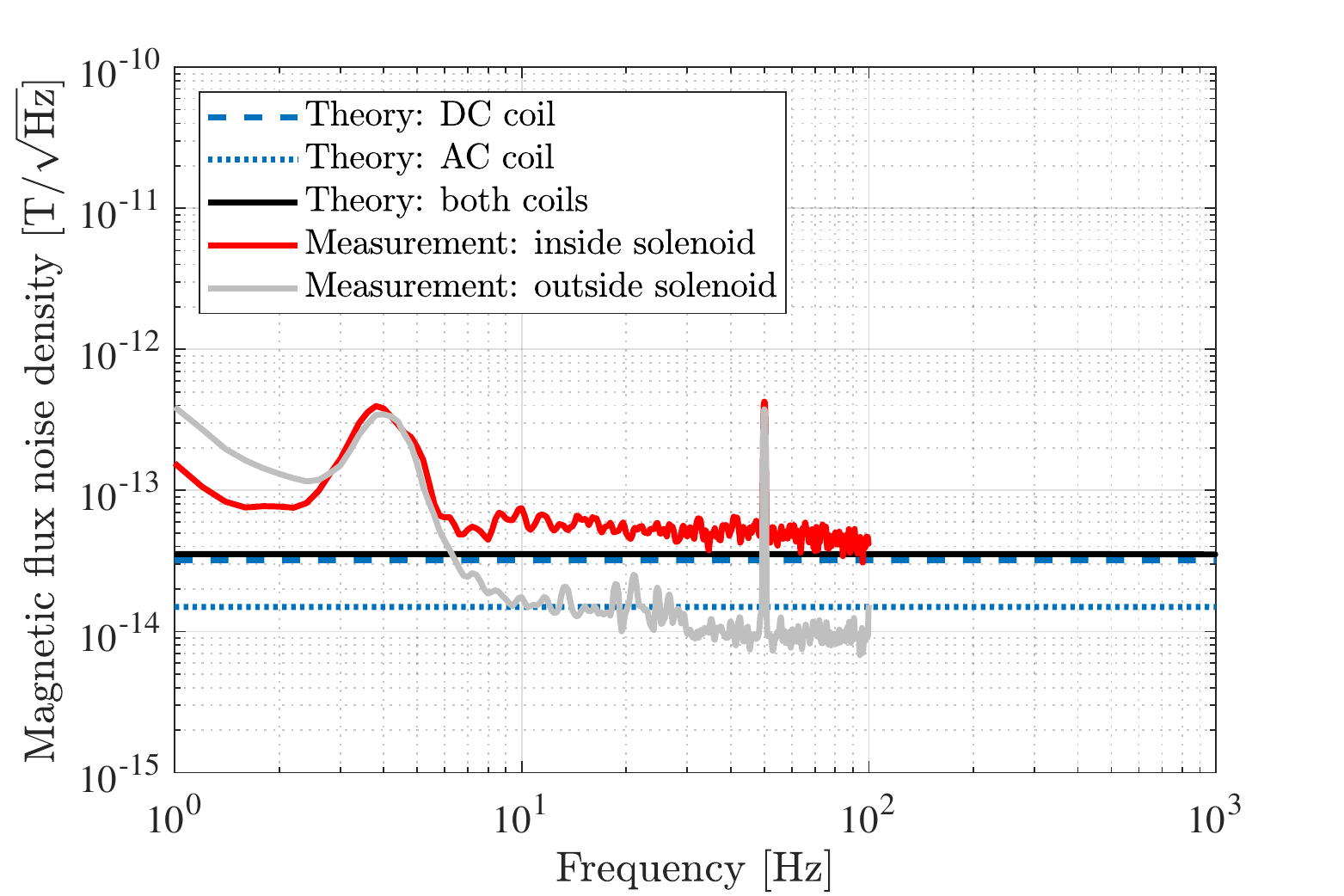}
	\caption{Calculated and measured magnetic flux noise density inside the solenoid. The measurements were performed with an optically pumped magnetometer and confirm the increased noise floor within the unconnected coil (red solid line) in comparison to a reference measurement outside the coil (gray solid line). Under the given circumstances that the detectivity of the utilized sensor is only slightly below the inherent noise of the coil and impairs the measurement accordingly, the measurement results can be regarded as verification of the model calculations.}
	\label{fig:thermal_noise}
\end{figure}

According to the fluctuation–dissipation theorem (FDT) every loss mechanism corresponds with fluctuations. In the particular case of a loss-representing electrical resistance the FDT simplifies to the well-known \textit{Johnson–Nyquist} theorem \cite{Joh28,Nyq28} that relates the voltage fluctuations, i.e. the voltage noise density ${V_{\mathrm{n}} = \sqrt{4 k_{\mathrm{B}} T R}}$ across the resistance $R$ with the temperature $T$ (in the following room temperature with $T = \SI{290}{K}$). The term $k_{\mathrm{B}}$ represents the \textit{Boltzmann} constant.

The coils under investigation exhibit two loss mechanisms as discussed above which are considered by the two resistances $R_{\mathrm{w}}$ and $R_{\mathrm{p}}$ in the equivalent circuit (Fig.~\ref{fig:equivalent_circuit}). Accordingly, there are also two voltage noise sources which can be described by the voltage noise densities
\begin{align}
	V_{\mathrm{nw}} = \sqrt{4 k_{\mathrm{B}} T R_{\mathrm{w}}}
\end{align}
and
\begin{align}
	V_{\mathrm{np}} = \sqrt{4 k_{\mathrm{B}} T R_{\mathrm{p}}}.
\end{align}
As shown above, the capacitance $C$ of the equivalent circuit can be neglected in the relevant frequency regime below \SI{10}{kHz}. Thus, based on the superposition principle, two simple expressions for the corresponding current noise densities through the coil can be derived that are given by
\begin{align}
	I_{\mathrm{nw}} = \frac{V_{\mathrm{nw}}}{|R_{\mathrm{w}} + R_{\mathrm{p}} + j \omega L|} = \sqrt{\frac{4 k_{\mathrm{B}} T R_{\mathrm{w}}}{(R_{\mathrm{w}} + R_{\mathrm{p}})^2 + (\omega L)^2}}
\end{align}
and
\begin{align}
	I_{\mathrm{np}} = \frac{V_{\mathrm{np}}}{|R_{\mathrm{w}} + R_{\mathrm{p}} + j \omega L|} = \sqrt{\frac{4 k_{\mathrm{B}} T R_{\mathrm{p}}}{(R_{\mathrm{w}} + R_{\mathrm{p}})^2 + (\omega L)^2}}.
\end{align}
With the magnetic sensitivity $S$, which relates a current through the coil with a magnetic flux density within the coil, the magnetic flux noise density is given by
\begin{align}
	B_{\mathrm{n}} &= S \cdot \sqrt{I_{\mathrm{nw}}^2 + I_{\mathrm{np}}^2}\\
	&= S \cdot \sqrt{\frac{4 k_{\mathrm{B}} T (R_{\mathrm{w}}+R_{\mathrm{p}})}{(R_{\mathrm{w}} + R_{\mathrm{p}})^2 + (\omega L)^2}}
\end{align}
when the two noise sources are assumed as statistically independent. Thus, the noise increases proportionally with the sensitivity of the coil. With the (measured) values as listed in Tab.~\ref{tab:coil_parameters} virtually frequency-independent values as low as ${B_{\mathrm{n}}^{\mathrm{(DC~coil)}} = 32~\mathrm{fT}/\sqrt{\mathrm{Hz}}}$ and ${B_{\mathrm{n}}^{\mathrm{(AC~coil)}} = 15~\mathrm{fT}/\sqrt{\mathrm{Hz}}}$ result (dashed and dotted lines in Fig.~\ref{fig:thermal_noise}). Due to the rigid connection of both coils, these values cannot be verified separately. Instead, only the sum of the noise contributions of both coils
\begin{align}
	B_{\mathrm{n}}^{\mathrm{total}} = \sqrt{\left(B_{\mathrm{n}}^{\mathrm{(DC~coil)}}\right)^2 + \left(B_{\mathrm{n}}^{\mathrm{(AC~coil)}}\right)^2}
	\label{eq:Bntotal}
\end{align}
can be measured, whereby it can be assumed that both noise contributions are statistically independent of each other. Based on Eq.~\eqref{eq:Bntotal}, a total magnetic flux noise density as low as $35~\mathrm{fT}/\sqrt{\mathrm{Hz}}$ can be expected (black solid line in Fig.~\ref{fig:thermal_noise}).

To verify this value, two magnetic noise measurements were performed with an optically pumped magnetometer (\textit{QZFM} from \textit{QuSpin} with a bandwidth of \SI{100}{Hz}) inside a magnetic shielding chamber (\textit{Vacuumschmelze GmbH \& Co. KG}, \cite[p. 117]{Jah13}). The measurement results are shown in Fig.~\ref{fig:thermal_noise} and basically confirm a magnetic flux noise density within the unconnected coil of several tens of $\mathrm{fT}/\sqrt{\mathrm{Hz}}$ (red solid line) in the frequency range between \SI{10}{Hz} and \SI{100}{Hz}. It is true that the measured values are slightly increased compared to the previously calculated noise. However, the increase in the noise level compared to a reference measurement outside the coil (gray solid line) is clearly visible. In addition, the intrinsic noise of the utilized OPM is also in the range of about $10~\mathrm{fT}/\sqrt{\mathrm{Hz}}$, which also affects the result of the noise measurement within the coil. Below \SI{10}{Hz} the inherent noise of the sensor and additional disturbances are even higher. Nevertheless, these results can be regarded as verification of the noise calculation and show in particular that the inherent magnetic noise of such a coil is largely negligible, at least when utilized with sensors based on the magnetostrictive effect.

\section{Magnetic Noise due to Current Noise of Extrinsic Current Sources}

\begin{figure*}[t!]
\captionsetup[subfigure]{justification=centering}
	\begin{subfigure}[t]{0.33\textwidth}
		\centering
		\includegraphics[width=1\linewidth]{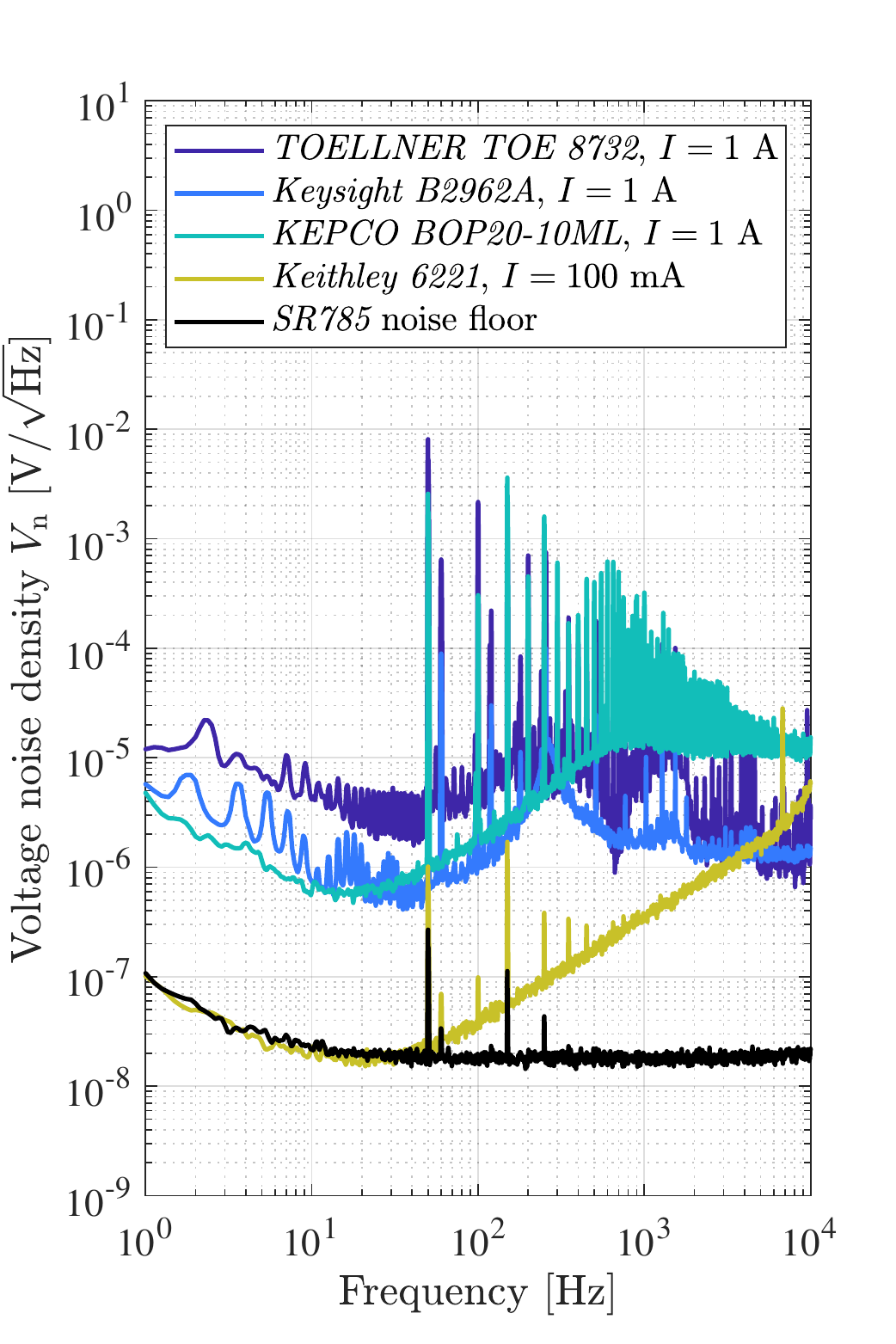}
		\caption{Voltage noise density across the coil}
		\label{fig:noise_commercial_sources_voltage}
	\end{subfigure}
	\begin{subfigure}[t]{0.33\textwidth}
		\centering
		\includegraphics[width=1\linewidth]{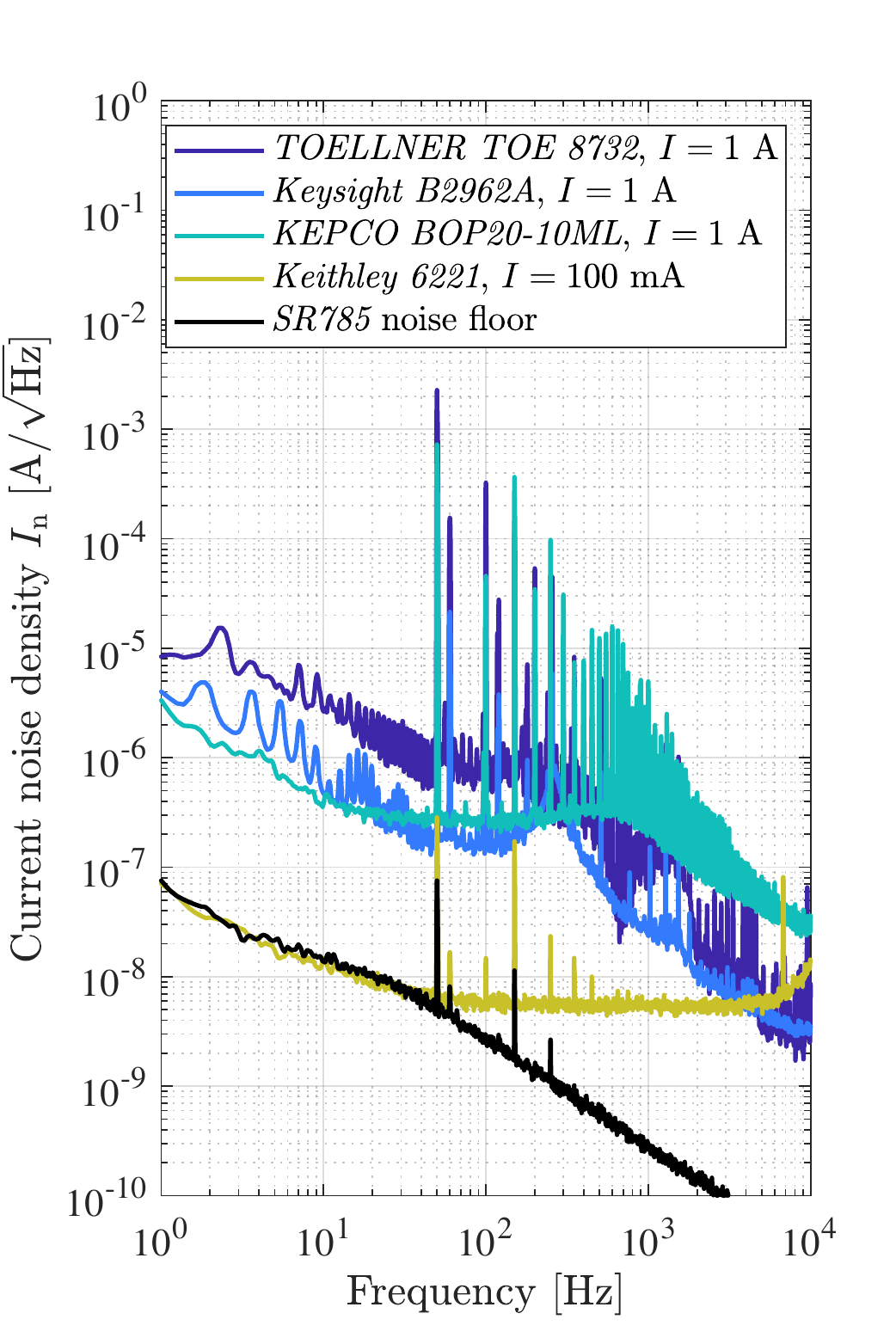}
		\caption{Current noise density through the coil}
		\label{fig:noise_commercial_sources_current}
	\end{subfigure}
	\begin{subfigure}[t]{0.33\textwidth}
		\centering
		\includegraphics[width=1\linewidth]{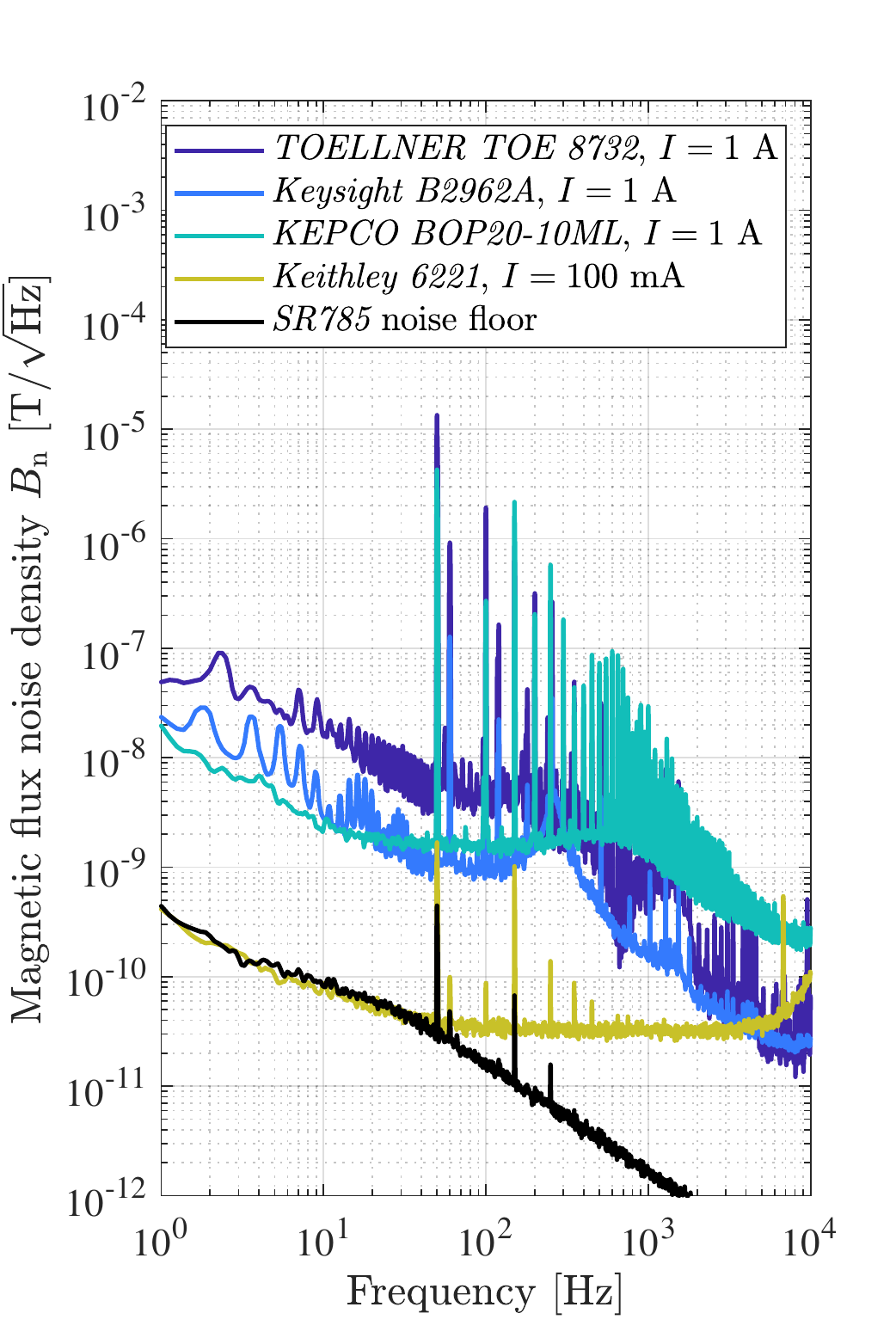}
		\caption{Magnetic flux noise density within the coil}
		\label{fig:noise_commercial_sources_magneticflux}
	\end{subfigure}
	\caption{Measurements of the magnetic flux noise density in the DC coil (\subref{fig:noise_commercial_sources_magneticflux}), by means of electrical measurements of the voltage noise density across the DC coil (\subref{fig:noise_commercial_sources_voltage}), and corresponding calculations of the current noise density through the coil (\subref{fig:noise_commercial_sources_current}) for various commercial current sources. The black line represents the noise limit of the measurement system, i.e. the noise floor of the utilized signal analyzer \textit{SR785} from \textit{Stanford Research Systems}.}
	\label{fig:noise_commercial_sources}
\end{figure*}

The inherent noise of the coil with low-frequency values in the range of a few tens of $\mathrm{fT}/\sqrt{\mathrm{Hz}}$ has turned out to be very low, i.e. in the same order of magnitude as the inherent noise level of state-of-the-art optically pumped magnetometers. If the coil is fed by an external current source to generate magnetic fields, it must be assumed that additional noise will occur. In the following, the additional noise of several commercial sources is characterized. However, since sensitive magnetic field sensors are generally saturated by the generated magnetic fields, the characterization is performed electrically instead.

For these characterizations, the magnetically shielded DC coil is supplied with relatively high currents while the voltage noise density across the DC coil $V_{\mathrm{n}}$ is measured with a dynamic signal analyzer \textit{SR785} from \textit{Stanford Research Systems}. With the previously determined electrical impedance of the coil, both the current noise density through the coil
\begin{align}
	I_{\mathrm{n}} = \frac{V_{\mathrm{n}}}{|Z_{\mathrm{coil}}|}
\end{align}
and the magnetic flux noise density within the coil
\begin{align}
	B_{\mathrm{n}} = S \cdot I_{\mathrm{n}} = S \cdot \frac{V_{\mathrm{n}}}{|Z_{\mathrm{coil}}|}
	\label{eq:Bn}
\end{align}
can be calculated, where the sensitivity $S$ is again utilized for the conversion between current and magnetic flux density. As before, the magnetic noise is directly proportional to the magnetic sensitivity of the coil. The results shown in Fig.~\ref{fig:noise_commercial_sources} reveal a significant increase in magnetic flux noise to values, depending on the frequency, in the $\mathrm{nT}/\sqrt{\mathrm{Hz}}$ regime if the coil is fed with a relatively high current of \SI{1}{A}. Only in case the coil is fed by a \textit{Keithley 6221} current source whose output current is limited to \SI{100}{mA} the noise floor is in the range of some tens of $\mathrm{pT}/\sqrt{\mathrm{Hz}}$. Although it can be seen that its noise increases with low frequencies, this is due to the noise limit of the measurement system, i.e. the inherent noise of the signal analyzer (black line).

\section{Analysis \& Conclusion}

We reported on the equivalent magnetic flux noise density within a pair of two coils that are typically used for the purpose of research and development of magnetic field sensors, e.g. sensors based on the magnetostrictive effect. Besides an analysis of the magnetic sensitivity and the electrical impedance of both coils, the focus was on the investigation of the coils noise behavior. Based on the losses within the coil, which are taken into account by corresponding resistors in the equivalent circuit, the inherent noise of the coil could be calculated and metrologically verified. In the low-frequency range, however, this magnetic flux density is only in the range of a few tens of $\mathrm{fT}/\sqrt{\mathrm{Hz}}$.

The DC coil is used for research and development purposes to generate static bias flux densities $B$, typically requiring values up to ${B = \SI{5}{mT}}$ depending on the magnetic material \cite{Lia20}. With a sensitivity of the investigated coil of ${S = \SI{5.95}{mT/A}}$ this value corresponds with a current of ${I = \SI{0.84}{A}}$. However, measurements have shown that the additional current noise $I_{\mathrm{n}}$ of commercial current sources then leads to a significant increase of the magnetic flux noise density $B_{\mathrm{n}}$ typically in the range of some $\mathrm{nT}/\sqrt{\mathrm{Hz}}$ (depending on the actual source and the frequency). Such values are obviously too high for application with magnetic field sensors with a low-frequency inherent noise below ${100~\mathrm{pT}/\sqrt{\mathrm{Hz}}}$.

In addition, it has been shown that a \textit{Keithley 6221} current source with a maximum current of ${I = \SI{100}{mA}}$ results in a significant lower magnetic flux noise density. Apart from the fact that the resulting noise is still well above ${B_{\mathrm{n}} = 10~\mathrm{pT}/\sqrt{\mathrm{Hz}}}$, one could come up with the idea to increase the magnetic sensitivity $S$ to solve the problem. This would result in lower current $I$ to generate a required magnetic bias flux density $B$. However, based on Eq.~\eqref{eq:S} and Eq.~\eqref{eq:Bn}, the magnetic-flux-density-to-magnetic-flux-noise-density ratio can be expressed as 
\begin{align}
	\frac{B}{B_{\mathrm{n}}} = \frac{S\cdot I}{S \cdot I_{\mathrm{n}}} = \frac{I}{I_{\mathrm{n}}}
\end{align}
which is obviously independent of the magnetic sensitivity $S$ and which is only determined by the current-to-current-noise ratio of the current source. For exemplary values of ${B = \SI{5}{mT}}$ and ${B_{\mathrm{n}} = 1~\mathrm{pT}/\sqrt{\mathrm{Hz}}}$ a current-to-current-noise ratio of 
\begin{align}
	\frac{I}{I_{\mathrm{n}}} = \frac{B}{B_{\mathrm{n}}} = \frac{\SI{5}{mT}}{1~\mathrm{pT}/\sqrt{\mathrm{Hz}}} = 500 \cdot 10^9~\sqrt{\mathrm{Hz}}
\end{align}
or
\begin{align}
	10 \cdot \log_{10} \left( \frac{\left( \frac{I}{I_{\mathrm{n}}} \right)^2}{\SI{1}{Hz}} \right)~\mathrm{dBHz} = 194~\mathrm{dBHz},
\end{align}
respectively, would be required. Although, to the best of our knowledge, no commercial current sources are available that reach that value, literature reports on ultra low-noise semiconductor current sources with achieved values as high as \SI{190}{dBHz} \cite{Cio98}, \SI{135}{dBHz} \cite{Lin04}, \SI{128}{dBHz} \cite{Eri08}, \SI{122}{dBHz} \cite{Tal11}, and \SI{172}{dBHz} \cite{Sca14}, each at a frequency of \SI{1}{Hz}. In addition, it must be mentioned that these current sources are usually limited to output currents of a few tens of $\mathrm{mA}$, so that a coil with a much higher magnetic sensitivity $S$ would still be required.

\section*{Acknowledgment}
This work was supported by the German Research Foundation (Deutsche Forschungsgemeinschaft, DFG) through the Collaborative Research Centre CRC 1261 \textit{Magnetoelectric Sensors: From Composite Materials to Biomagnetic Diagnostics}. In addition, the authors would like to thank Eric Elzenheimer for supporting the magnetic noise measurements with an optically pumped magnetometer.

\ifCLASSOPTIONcaptionsoff
  \newpage
\fi




\bibliographystyle{IEEEtran}
\bibliography{mybibfile}

\begin{thebibliography}{10}
\providecommand{\url}[1]{#1}
\csname url@samestyle\endcsname
\providecommand{\newblock}{\relax}
\providecommand{\bibinfo}[2]{#2}
\providecommand{\BIBentrySTDinterwordspacing}{\spaceskip=0pt\relax}
\providecommand{\BIBentryALTinterwordstretchfactor}{4}
\providecommand{\BIBentryALTinterwordspacing}{\spaceskip=\fontdimen2\font plus
\BIBentryALTinterwordstretchfactor\fontdimen3\font minus
  \fontdimen4\font\relax}
\providecommand{\BIBforeignlanguage}[2]{{%
\expandafter\ifx\csname l@#1\endcsname\relax
\typeout{** WARNING: IEEEtran.bst: No hyphenation pattern has been}%
\typeout{** loaded for the language `#1'. Using the pattern for}%
\typeout{** the default language instead.}%
\else
\language=\csname l@#1\endcsname
\fi
#2}}
\providecommand{\BIBdecl}{\relax}
\BIBdecl

\bibitem{Hor15}
P.~Horowitz and W.~Hill, \emph{{The Art of Electronics}}.\hskip 1em plus 0.5em
  minus 0.4em\relax New York City, New York, USA: Cambridge University Press,
  2015.

\bibitem{Lag12}
E.~Lage, C.~Kirchhof, V.~Hrkac, L.~Kienle, R.~Jahns, R.~Kn{\"{o}}chel,
  E.~Quandt, and D.~Meyners, ``{Exchange biasing of magnetoelectric
  composites},'' \emph{Nature Materials}, vol.~11, no.~6, pp. 523--529, Jun.
  2012.

\bibitem{Dur20}
P.~Durdaut, E.~Rubiola, J.-M. Friedt, C.~M{\"{u}}ller, B.~Spetzler,
  C.~Kirchhof, D.~Meyners, E.~Quandt, F.~Faupel, J.~McCord, R.~Kn{\"{o}}chel,
  and M.~H{\"{o}}ft, ``{Phase Sensitivity and Phase Noise of Cantilever-Type
  Magnetoelastic Sensors Based on the $\Delta$E Effect},'' {Preprint
  arXiv:2003.01085 (physics.ins-det), March 2020}.

\bibitem{Sch20}
V.~Schell, C.~M{\"{u}}ller, P.~Durdaut, A.~Kittmann, L.~Thorm{\"{a}}hlen,
  F.~Lofink, D.~Meyners, M.~H{\"{o}}ft, J.~McCord, and E.~Quandt, ``{Magnetic
  anisotropy controlled FeCoSiB thin films for surface acoustic wave magnetic
  field sensors},'' \emph{Applied Physics Letters}, vol. 116, no.~7, pp.
  073\,503 1--5, Feb. 2020.

\bibitem{Ser04}
R.~A. Serway and J.~W. Jewett, \emph{{Physics for Scientists and Engineers}},
  6th~ed.\hskip 1em plus 0.5em minus 0.4em\relax Boston, Massachusetts, USA:
  Brooks Cole, 2004.

\bibitem{Bow97}
C.~Bowick, \emph{RF circuit design}.\hskip 1em plus 0.5em minus 0.4em\relax
  Burlington, Massachusetts, USA: Newnes, 1997.

\bibitem{Lot92}
A.~Lotfi, P.~Gradzki, and F.~Lee, ``{Proximity effects in coils for high
  frequency power applications},'' \emph{IEEE Transactions on Magnetics},
  vol.~28, no.~5, pp. 2169--2171, Sep. 1992.

\bibitem{Dow66}
P.~L. Dowell, ``{Effects of eddy currents in transformer windings},''
  \emph{Proceedings of the Institution of Electrical Engineers}, vol. 113,
  no.~8, pp. 1387--1394, 1966.

\bibitem{Joh28}
J.~B. Johnson, ``{Thermal Agitation of Electricity in Conductors},''
  \emph{Physical Review}, vol.~32, no.~1, pp. 97--109, Jul. 1928.

\bibitem{Nyq28}
H.~Nyquist, ``{Thermal Agitation of Electric Charge in Conductors},''
  \emph{Physical Review}, vol.~32, no.~1, pp. 110--113, Jul. 1928.

\bibitem{Jah13}
\BIBentryALTinterwordspacing
R.~Jahns, ``{Untersuchung und Optimierung von Empfindlichkeit und
  Rauschverhalten magnetoelektrischer Sensoren},'' Ph.D. dissertation, Kiel
  University, 2013. [Online]. Available:
  \url{https://macau.uni-kiel.de/receive/diss_mods_00012442}
\BIBentrySTDinterwordspacing

\bibitem{Lia20}
X.~Liang, C.~Dong, H.~Chen, J.~Wang, Y.~Wei, M.~Zaeimbashi, Y.~He,
  A.~Matyushov, C.~Sun, and N.~Sun, ``{A Review of Thin-Film Magnetoelastic
  Materials for Magnetoelectric Applications},'' \emph{Sensors}, vol.~20,
  no.~5, p. 1532, Mar. 2020.

\bibitem{Cio98}
C.~Ciofi, R.~Giannetti, V.~Dattilo, and B.~Neri, ``{Ultra Low-Noise Current
  Sources},'' \emph{IEEE Transactions on Instrumentation and Measurement},
  vol.~47, no.~1, pp. 78--81, 1998.

\bibitem{Lin04}
S.~Linzen, T.~L. Robertson, T.~Hime, B.~L.~T. Plourde, P.~A. Reichardt, and
  J.~Clarke, ``{Low-noise computer-controlled current source for quantum
  coherence experiments},'' \emph{Review of Scientific Instruments}, vol.~75,
  no.~8, pp. 2541--2544, Aug. 2004.

\bibitem{Eri08}
C.~J. Erickson, M.~{Van Zijll}, G.~Doermann, and D.~S. Durfee, ``{An ultrahigh
  stability, low-noise laser current driver with digital control},''
  \emph{Review of Scientific Instruments}, vol.~79, no.~7, pp. 073\,107 1--8,
  Jul. 2008.

\bibitem{Tal11}
D.~Talukdar, R.~K. Chakraborty, S.~Bose, and K.~K. Bardhan, ``{Low noise
  constant current source for bias dependent noise measurements},''
  \emph{Review of Scientific Instruments}, vol.~82, no.~1, pp. 013\,906 1--6,
  Jan. 2011.

\bibitem{Sca14}
G.~Scandurra, G.~Cannat{\`{a}}, G.~Giusi, and C.~Ciofi, ``{Programmable, very
  low noise current source},'' \emph{Review of Scientific Instruments},
  vol.~85, no.~12, pp. 125\,109 1--10, Dec. 2014.

\end{thebibliography}


%






\vfill


\end{document}